\documentclass[aps,prl,reprint,longbibliography,superscriptaddress,nofootinbib]{revtex4-2}
\usepackage[pdftex]{graphicx}
\usepackage{bm}
\usepackage[pdftex,colorlinks=false,citecolor=black,urlcolor=black,linkcolor=black]{hyperref}
\usepackage{braket}
\usepackage[dvipsnames]{xcolor}
\usepackage[version=3]{mhchem}
\usepackage{float}
\emergencystretch=\maxdimen
\usepackage[locale = US]{siunitx}
\DeclareSIUnit\G{G}
\DeclareSIUnit\kb{\textit{k}_{\hspace{1pt}B}}

\usepackage{amssymb}
\usepackage{amsmath}
\usepackage{old-arrows}

\newcommand{\updownarrows}{\uparrow\joinrel\downarrow}

\begin{document}

\title{Programmable Assembly of Ground State Fermionic Tweezer Arrays}

\author{Naman Jain}
\affiliation{Max Planck Institute of Quantum Optics, Hans-Kopfermann-Str. 1, Garching 85748, Germany}

\author{Jin Zhang}
\affiliation{Max Planck Institute of Quantum Optics, Hans-Kopfermann-Str. 1, Garching 85748, Germany}

\author{Marcus Culemann}
\affiliation{Max Planck Institute of Quantum Optics, Hans-Kopfermann-Str. 1, Garching 85748, Germany}

\author{Philipp M. Preiss}
\email{philipp.preiss@mpq.mpg.de}
\affiliation{Max Planck Institute of Quantum Optics, Hans-Kopfermann-Str. 1, Garching 85748, Germany}
\affiliation{Munich Center for Quantum Science and Technology (MCQST), Schellingstr. 4, München 80799, Germany}

\date{\today}

\begin{abstract}

We demonstrate deterministic preparation of arbitrary two-component product states of fermionic $^6$Li atoms in an 8$\times$8 optical tweezer array, achieving motional ground-state fidelities above 98.5\%. Leveraging the large differential magnetic moments for spin-resolution, with parallelized site- and number-resolved control, our approach addresses key challenges for low-entropy quantum state engineering. Combined with high-fidelity spin-, site-, and density-resolved readout within a single \qty{20}{\us} exposure, and \qty{3}{\s} experimental cycles, these advances establish a fast, scalable, and programmable architecture for fermionic quantum simulation.

\end{abstract}

%\keywords{Quantum gas assembly, Ultracold Atoms, Tweezer Arrays, Fermionic Gas, Deterministic state preparation, spin and site resolution with Li6, scalable scheme}
 
\maketitle
%%%%%%%%%%%%%%Central Text%%%%%%%%%%%
%%%%% \section{Introduction} \label{sec:intro}
Exercising precise control over microscopic quantum systems is a central theme in modern physics, underpinning advances across quantum simulation, information science, and metrology~\cite{quantumsimulation_review, QI_Rydbergs, clocks_review}. Neutral atom platforms provide a powerful approach to this endeavor, as demonstrated by rapid progress in their deterministic initialization and high-fidelity detection~\cite{bloch_gross_review, QCwithNeutralAtoms,bakrreview_qgm}. While fast single-atom preparation in optical tweezers via light-assisted collisions has enabled defect-free assembly of atomic arrays~\cite{KaufmanHighFillingTweezer,Bluvstein2024, SQM_Tweezers}, initializing many atoms into a single programmable quantum state remains an experimental challenge. Achieving full control over internal and external degrees of freedom is particularly important for fermions, where strongly correlated Hubbard physics arises from the interplay between spin and charge correlations~\cite{fewfermionprep, Xu2025}. Arbitrary experimental control over many fermionic atoms at the level of single quantum states would allow the assembly of such Hubbard systems, with interesting applications in the study of out-of-equilibrium entanglement dynamics, spin transport, and quantum thermalization~\cite{QMBSoutofequilibrium_Eisert, quenchdynamicsreview, hilbertspacefragmentation}.

Fermionic quantum simulators based on quantum gas microscopes have uncovered rich itinerant many-body physics, including antiferromagnetic correlations, quasiparticle dynamics, and pseudogap signatures--through a top-down approach using bulk loading into optical lattices~\cite{Lithium_AFM_order, greiner_pseudogap, fermionpairing_zwierlein}. In practice, this requires sophisticated cooling and entropy redistribution schemes, which restrict the accessible configurations and often result in experimental cycle times of tens of seconds~\cite{qgm_review_gross_bakr}. Complementary bottom-up approaches using few-fermion preparation in tweezers have enabled studies of quantum statistics, Laughlin states, and tunnel-coupled arrays up to eight sites~\cite{manyparticleinterference,Jochim_Laughlin, Bakr_2D}. These successes motivate architectures that merge the strengths of both approaches~\cite{kaufman_quantum_walk}, realizing high repetition rates, low-entropy initialization with local spin control, and high-fidelity readout. 

\begin{figure*}[ht]
	\centering\
	\includegraphics[width=\linewidth]{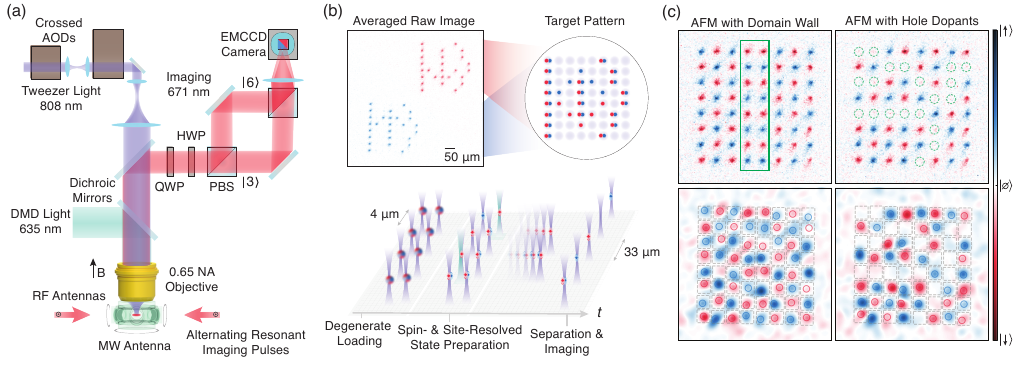} %
	\caption{\textbf{Arbitrary product state preparation and high-fidelity detection of fermionic atoms. (a) }Schematic of the setup. A high-numerical aperture objective creates 2D tweezer arrays via orthogonal AODs, with superimposed DMD potentials, while collecting \qty{671}{\nm} fluorescence for detection. Spin information is encoded in orthogonally polarized fluorescence photons and separated via polarization optics onto distinct EMCCD regions. \textbf{(b)} Simultaneous spin detection of $^6$Li ($\ket{3}$: red, $\ket{6}$: blue), shown in an averaged raw image. Example pattern demonstrates arbitrary product state preparation on an 8$\times$8 array; target pattern is shown on the right. The lower panel highlights the programmability of our platform using tweezers (purple) and DMD potentials (green, meshed grid). \textbf{(c)} Representative assembled states for Hubbard physics with overlapped spin sectors. Left: classical anti-ferromagnet (AFM) with a vertical domain wall (green). Right: hole-doped AFM. Top: averaged raw images; bottom: processed single-shot realizations, in arrays loaded at~\qty{4}{\um}, and imaged at~\qty{33}{\um} separation.}
	\label{fig:Image1_Teaser}
\end{figure*}
Here, we demonstrate such an approach by assembling and detecting 2D fermionic atom arrays with arbitrary density and spin distributions (Fig.~\ref{fig:Image1_Teaser}), achieving complete control over the relevant internal and motional degrees of freedom within \qty{3}{\s} cycle times. This level of control emerges from several advances developed in this Letter. We establish a robust loading protocol that draws degenerate Fermi samples from a cold reservoir into a tweezer array without being limited by reservoir size. Local evaporation in the tweezers enables deterministic preparation of singlet pairs in the 3D motional ground state of each trap, as first demonstrated in~\cite{fewfermionprep}. To achieve site-resolved spin control in $^6$Li, we exploit the Zeeman structure to tune the differential magnetic moment between the two lowest hyperfine states, and combine it with locally tailored optical gradients generated by a digital micromirror device (DMD), providing parallel spin selectivity across the array. This approach enables the initialization of programmable spin-charge configurations with controlled defects, including classical antiferromagnetic patterns [Fig.~\ref{fig:Image1_Teaser}(c)]. We probe these on-demand product states with a single-exposure imaging scheme that resolves both spin and density within \qty{20}{\us} by scattering on stretched states, extending the scheme from~\cite{bergschneider1} (also see Ref.~\cite{HQA_imaging}). Fast internal-state control is achieved through radio-frequency (RF) and microwave (MW) antennas that deliver the highest reported Rabi rates on the relevant transitions~\cite{Roati_antenna}. These capabilities also position $^6$Li as a promising candidate for fermionic quantum computation.

Together, these advances enable programmable, low-entropy initial states with intrinsic fermionic statistics, granting access to computationally challenging many-body physics and emerging architectures for fermionic quantum processing~\cite{fermioniccomputing_cuadra, errorcorrection_ott, faulttolerantQC_schukert}. Combined with dynamics in lattices or tunnel-coupled tweezers, they open new opportunities for exploring strongly correlated phases and nonequilibrium behavior. The ability to assemble large product states atom by atom further facilitates protocols that require structured input states, such as Hamiltonian learning, as well as studies of precisely controlled impurity-, disorder-, and frustration-induced phenomena~\cite{hamiltonianlearning_advantageofquantum, impurityeffects_review}.

\begin{figure}[ht]
	\includegraphics[width=82mm]{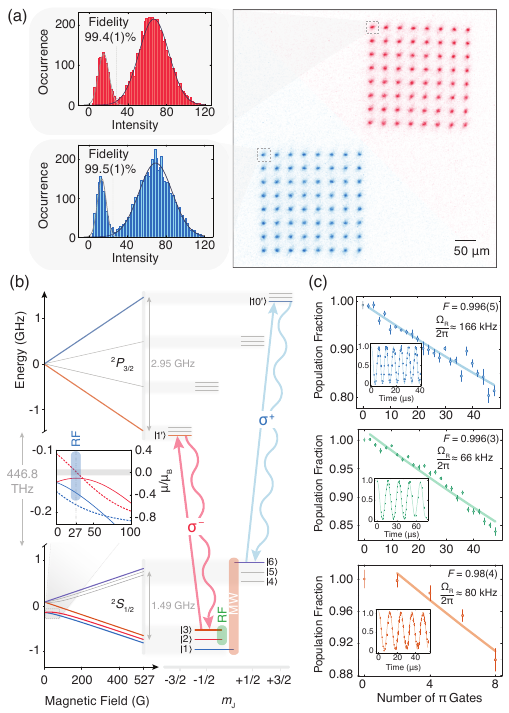} %
	\caption{\textbf{Rapid single-exposure quantum state readout of $^6$Li. (a)} Averaged single-exposure image of a fully filled array with representative histograms demonstrating high-fidelity 0-1 atom discrimination for both spins. \textbf{(b)} The addressed cycling $D2$ transitions, $\ket{6}\to\ket{10'}$ and $\ket{3}\to\ket{1'}$ enable spin-resolved detection. Zeeman energies (solid) and magnetic moments $\mu$ (\textit{inset}, dashed) for $\ket{1}$ (blue) and $\ket{2}$ (red), leveraged for state-dependent potentials. Shaded bands denote the RF and MW antenna-driven transitions. \textbf{(c)} Repeated $\pi$-pulse fidelity ($F$) for the state transfers: $\ket{1} - \ket{2}$ (\qty{24.3}{\MHz}, blue) for spin-resolved preparation at \qty{27}{\G}, $\ket{2}-\ket{3}$ (\qty{84.6}{\MHz}, green), and $\ket{1}-\ket{6}$ (\qty{1.64}{\GHz}, orange) utilized for detection at \qty{527}{\G}. Insets show the measured coherent oscillations with Rabi frequency $\Omega_R$.}
	\label{fig:Image2_SpinResolvedImaging}
\end{figure}

\textit{Fast spin-resolved imaging}--- We begin by introducing a spin- and density-resolved detection scheme, essential for probing state preparation in atomic arrays. For $^6$Li, spin-resolved imaging is challenging due to its low mass and small fine and hyperfine constants. Existing approaches often rely on mapping internal states onto auxiliary external atomic degrees of freedom, and typically require deep pinning lattices with long (hundreds of ms) exposures~\cite{greiner_preiss_imaging, lithium_imaging, zwierlein_imaging, Bakr_2D}. Building on the free-space approach of~\cite{bergschneider1}, we instead obtain spin-resolution with single-atom sensitivity in a single \unit{\us}-scale exposure by encoding the spin information in the fluorescence polarization, adopting the strategy from~\cite{HQA_imaging}. We resonantly drive closed $D2$ ($\sigma^\pm$) transitions from the stretched states $\ket{6}$ and $\ket{3}$, to which the science states ($\ket{1}$ and $\ket{2}$) are transferred before detection. At \qty{527}{\G}, these lines are split by \qty{1.46}{\GHz}, resulting in negligible off-resonant scattering. Using polarizing beam splitters in the imaging path, the collected fluorescence from the two orthogonal polarizations is directed onto distinct regions of an electron-multiplying charge-coupled device (EMCCD) camera, separated by $\approx \,$\qty{4.7}{\mm} [Fig.~\ref{fig:Image2_SpinResolvedImaging}(a)]. The key advantage of this method is that a single imaging pulse with two frequency components creates spatially separated images of the two atomic spin states without inherent loss of fluorescence photons.

Figure~\ref{fig:Image2_SpinResolvedImaging}(a) shows the spin-resolved image and histograms obtained for an 8$\times$8 tweezer array loaded with at most one particle per site and spin state. A \qty{19}{\micro\second} exposure yields $\approx \,$35 detected photons per atom, resulting in an imaging fidelity above 99.4\%. The complete detection sequence, including internal state transfers, realizes a total detection fidelity exceeding $98.5\%$ for both spins~\cite{Supplement}\nocite{Li2D_MOT,hqa,rf_phase,GRIMM200095,mjt,akali_earth_magnetic_moment, Su2025, Bakr_1D}. We implement state mappings within \qty{15}{\us} at \qty{527}{\G}. The $\ket{2}-\ket{3}$ transition is driven by a PCB-based RF antenna, which achieves $99.6(5) \%$ fidelity over a \qty{7}{\us} $\pi$ pulse [Fig.~\ref{fig:Image2_SpinResolvedImaging}(c)]. For the $\ket{1}-\ket{6}$ transition, we employ a single-loop MW antenna with an optimized wideband, uniform rate, smooth truncation (WURST) pulse~\cite{GOIA-WURST} to provide robustness against magnetic field noise. Although slow field drifts limit multipulse fidelity to $\approx \,98\%$, a single \qty{15}{\us} optimized $\pi$ pulse achieves $99.03(5)\%$ transfer fidelity~\cite{Supplement}. Additionally, we observe that $\ket{3}-\ket{6}$ atoms confined to the same site undergo rapid loss on a few-ms timescale due to spin-changing collisions. To suppress this, the tweezers are switched off prior to state transfers, followed immediately by the imaging pulses. This protocol enables the single-exposure, spin-resolved readout of the tweezer array.

\begin{figure}%[ht]
	\centering
	\includegraphics[width=82mm]{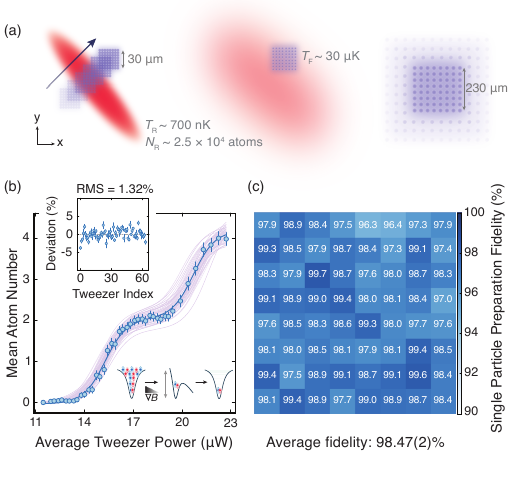} %
	\caption{\textbf{Deterministic singlet preparation in tweezer arrays. (a)} Homogeneous loading is achieved in the tweezer array by translating it across a cold $\ket{1}-\ket{2}$ reservoir formed by a \qty{1070}{\nm} ODT. Site occupations are detected after array expansion. \textbf{(b)} Typical spilling curves at $\nabla B \,$= \qty{21.5}{\G / \cm} for individual tweezers (light lines) closely follow the array-average trend (dark line), demonstrating power uniformity (\textit{inset}). The even-atom plateaus characterize strong two-component Fermi degeneracy. The schematic illustrates selective removal of excited motional states via spilling. \textbf{(c)} Heatmap depicting single-particle preparation fidelities for individual tweezers across the 8$\times$8 array with best tweezers reaching $99.7\%$.}
	\label{fig:Image3_deterministicprep}
\end{figure}

%%%%%%\textit{Singlet Preparation in Large Tweezer Arrays}:
\textit{Singlet preparation in programmable tweezer arrays}---Our approach to initialization of low-entropy fermionic product states begins with preparing a uniform ensemble of ground-state $\ket{1}-\ket{2}$ singlet pairs in an optical tweezer array, which serve as a starting point toward engineering structured spin-charge patterns. As shown in Fig.~\ref{fig:Image1_Teaser}(a), an 8$\times$8 array of optical tweezers [$\lambda_{\mathrm{T}}$= \qty{808}{\nm}, waist= \qty{560 \pm 30}{\nm}] is generated using orthogonal acousto-optic deflectors (AODs), providing continuous 2D positioning and depth control via multitone RF signals~\cite{Supplement}. Ground-state loading exploits Pauli suppression of density fluctuations at the bottom of a Fermi sea, as pioneered in~\cite{fewfermionprep}. To ensure this, we create degenerate samples using exponential enhancement of phase-space density through adiabatic overlap of traps with large volume ratios~\cite{reversiblebec}. A crossed optical dipole trap (ODT) containing an attractively interacting $\ket{1}-\ket{2}$ mixture at $T_{\mathrm{R}}\lesssim \,$\qty{700}{\nano\kelvin} serves as a reservoir for loading tweezers at \qty{50}{\micro\kelvin \cdot \kb} depth and $\approx \,$\qty{4}{\um} spacing to ensure independent loading [Fig.~\ref{fig:Image3_deterministicprep}(a)]. The tweezer-trapped sample ($T_{\mathrm{F}} \! \approx \,$\qty{30}{\micro\kelvin}) thermalizes with the reservoir, yielding high degeneracy $T_{\mathrm{R}}/T_{\mathrm{F}} \approx \,0.025$ and thus near-unity ground state filling. For arrays sizes comparable to the reservoir, spatial variations in the energy difference between tweezer depth and the reservoir chemical potential lead to inhomogeneous loading. We mitigate this by translating the entire array across the ODT over $50-$\qty{100}{\ms} at $\approx \,$\qty{0.4}{\um / \ms}, allowing thermalization near the reservoir center, while avoiding rethermalization at the edges. This results in uniform loading of tens of atoms per trap, and scales to larger arrays, with current limits set by imaging constraints. This approach opens up prospects of in-sequence Fermi gas replenishment using separate tweezer arrays for system and reservoir.
\begin{figure*}
	\centering
	\includegraphics[width=\linewidth]{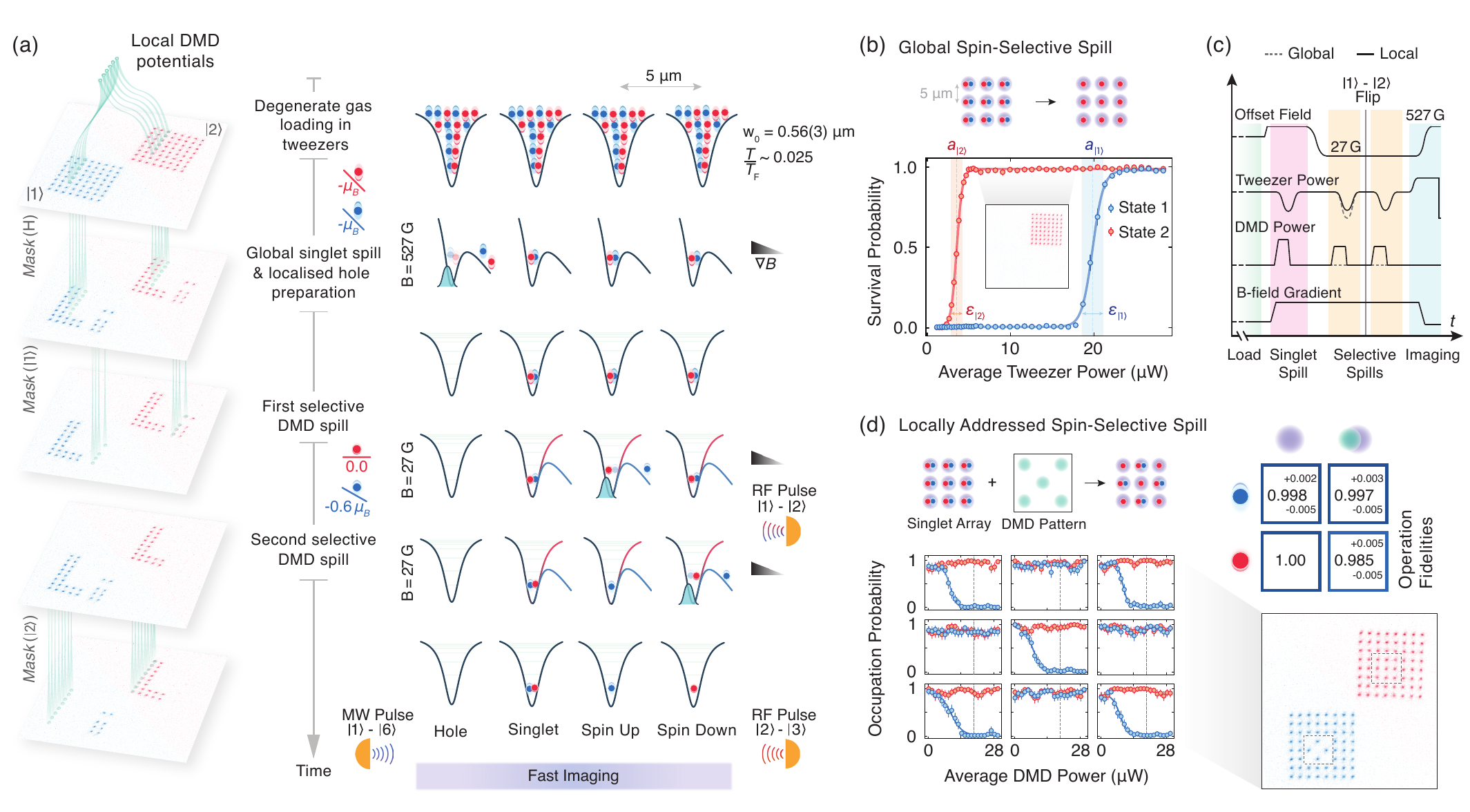} 
	\caption{\textbf{Arbitrary product state preparation. (a)} Schematic of the protocol for preparing a $\ket{2}-\ket{1}$ spin-patterned `\textit{L-i}'. Left: representative DMD potentials (showing only the first addressed column for clarity; green) with spin-resolved images. Right: axial potential schematic summarizing the concept. First, a global, spin-agnostic spill removes excited motional states, while a repulsive DMD mask, \textit{Mask}($H$), simultaneously creates site-resolved holes. A second DMD pattern \textit{Mask}($\ket{1}$) targets sites designated for $\ket{1}$ occupation, followed by a $\ket{1}-\ket{2}$ population swap, and a final pattern \textit{Mask}($\ket{2}$) that imprints the complementary spin structure, realizing arbitrary spin-charge configurations. \textbf{(b)} Measured spin-selective survival probabilities at fixed $\nabla B \,$= \qty{50}{\G / \cm}. Sigmoid fits yield centers $a_{\ket{1}}$=\qty{19.83\pm 0.03}{\uW}, $a_{\ket{2}}$=\qty{3.54\pm 0.02}{\uW}, and widths $\varepsilon_{\ket{1}}$= \qty{1.3 \pm 0.06}{\uW}, $\varepsilon_{\ket{2}}$= \qty{0.70 \pm 0.02}{\uW}. \textbf{(c)} Experimental sequence mapping key parameters. \textbf{(d)} Local spin-selective spilling using the DMD at fixed tweezer power ($\approx \,$\qty{16}{\uW}) and $\nabla B \,$= \qty{21.5}{\G / \cm}. Operational fidelities correspond to the probability of retaining (removing) a specific spin at addressed (nonaddressed) sites in the 3D ground state, evaluated at DMD power marked by the dashed line.}
	\label{fig:Image4_spindependentprep}
\end{figure*}

From degenerate samples, we create independent singlet pairs by tuning the magnetic field offset to \qty{527}{\G}, where the $\ket{1}-\ket{2}$ mixture is effectively noninteracting, and both states have magnetic moments $\approx -\mu_{\mathrm{B}}$ (Bohr magneton). Applying a field gradient with controlled reduction of tweezer depths, we selectively retain only the lowest bound states~\cite{fewfermionprep} [Fig.~\ref{fig:Image3_deterministicprep}(b)]. The trap-depths are ramped globally by stabilizing the laser power before the AODs, using Blackman-shaped pulses to suppress motional excitations. Depth homogeneity is achieved by iteratively tuning the $2N$ RF-tone amplitudes using atomic feedback, exploiting the steep power sensitivity between the zero- and two-atom plateaus in individual spilling curves~\cite{Bakr_2D}(see Fig.~\ref{fig:Image3_deterministicprep}). The resulting $< \! 1.4\%$ depth variation ensures uniform spilling across the 8$\times$8 array, with the operating point chosen such that residual imperfections manifest predominantly as an atom loss rather than overfilling, verified via atom-resolved counting in a magneto-optical trap~\cite{fewfermionprep,motcounting_oberthaler, Supplement}. We benchmark preparation fidelity by performing an additional spill after the sequence to remove any atoms in motional excited states, directly probing the 3D ground-state population~\cite{Supplement}. The single-particle ground-state fidelity averages $98.47(2)\%$, reaching \num{99.7(0.3:0.4)}\% in the best tweezers, with ground-state lifetimes of 17.5(1.0)~\unit{\s}. Using balanced 2D arrays, we also realize magnetic-gradient-free singlet preparation relying solely on gravity, and achieve similar array-averaged fidelities ($\approx \! 98.7(1)\%$), demonstrating the applicability of this approach to fermions with negligible magnetic moment~\cite{Supplement}. For experimental brevity, we use gradient-assisted spilling for all measurements presented here. These fidelities compare well to the state-of-the-art neutral-atom platforms employing in-trap cooling schemes~\cite{KaufmanSingleAtomTweezerCooling, SQM_Tweezers, Kaufman_YtTweezers, kaufman_quantum_walk}. Such initialization establishes the singlet array as a uniform, low-entropy baseline for programmable fermionic assembly. \\

%%%%%% \section{Parallelised Spin-Dependent Preparation} \label{sec:spindependentprep}
\textit{Parallelized spin-dependent preparation}---The generation of arbitrary product states further requires precise local spin control. In $^6$Li, this is complicated by the small fine structure splitting ($\approx \,$\qty{10}{\GHz}), which limits optical selectivity by inducing strong off-resonant scattering~\cite{SpinDependentPotential}. Instead, we harness the tunable magnetic moments of the internal states to realize lossless, spin-selective potentials~\cite{fewfermionprep}. In particular, at $\qty{27}{\G}$, $\ket{2}$ exhibits a vanishing magnetic moment, while $\ket{1}$ retains $\mu_{\ket{1}} \! \approx \! -0.6\mu_{\mathrm{B}}$ [see Fig.~\ref{fig:Image2_SpinResolvedImaging}(b)]. In a magnetic field gradient, this differential moment yields a spin-dependent potential slope, allowing selective spilling of $\ket{1}$ atoms while leaving $\ket{2}$ unperturbed~\cite{Jochim_few_to_many}. We demonstrate this concept by performing a global spin-selective spill on an 8$\times$8 array of $\ket{1}-\ket{2}$ singlet pairs [Fig.~\ref{fig:Image4_spindependentprep}(b)]. The measured ground-state survival probabilities show sharply separated extinction curves for the two spins, corresponding to a sixfold difference in tweezer depths for the bound state thresholds. We emphasize that this magnetically induced selectivity is very robust and does not rely on the fine-tuned tweezer depths required for singlet preparation. 

Building on this, we realize site-resolved control using programmable optical potentials projected by a DMD. Spatially incoherent \qty{635}{\nm} light is used to generate localized repulsive disks with a radius of $\approx \,$\qty{500}{\nm}, aligned to individual tweezers with $< \,$\qty{130}{\nm} error. By offsetting the DMD focal plane by a few micrometers, we impose controllable axial gradients on each tweezer. For typical tweezer spilling depths of $\approx \,$\qty{85}{\kHz}, the DMD can apply site-specific potentials of $\approx \,$\qty{500}{\kHz} at its focus--corresponding to optical gradients of $\approx \,$\qty{100}{\kHz / \um} on the atoms. Gray scaling over $\approx \,$113 mirrors per site provides $\approx \,$\qty{1}{\%} depth resolution, and hence of optical gradients. Crucially, the DMD-generated addressing patterns enable independent control over the effective depth of each tweezer, resulting in fully parallelized local control.

In the experiment, following singlet preparation, spin-selective atom removal is performed by combining a magnetic-field gradient, a global tweezer power ramp, and DMD-defined local potentials. The gradient weakens the effective confinement only for $\ket{1}$ atoms, whereas gradient-insensitive $\ket{2}$ atoms remain stably trapped [Fig.~\ref{fig:Image4_spindependentprep}(a)]. A localized, spin-independent DMD-induced optical gradient on selected sites then nudges only the $\ket{1}$ atoms across the bound-state threshold, leading to their clean removal. In a representative `cross' pattern [Fig.~\ref{fig:Image4_spindependentprep}(d)], addressed tweezers show \num{99.7(0.3:0.5)}\% removal fidelity for $\ket{1}$ and \num{98.5(0.5:0.5)}\% probability for $\ket{2}$ retention in the ground state. Nonaddressed sites retain $\ket{1}$ atoms with \num{99.8(0.2:0.5)}\% probability and show no detectable change in $\ket{2}$, compared to the initial ground-state sample. 

Equipped with site-resolved spin selectivity, we realize arbitrary product states in the 8$\times$8 array (Fig.~\ref{fig:Image1_Teaser}). As outlined in Fig.~\ref{fig:Image4_spindependentprep}(a), our protocol first employs a DMD mask (\textit{Mask}(H)) that defines the hole pattern during a spin-agnostic global spill, followed by a low-field spin-selective removal with \textit{Mask}($\ket{1}$) to define the $\ket{1}$ sector. A \qty{3.5}{\us} RF $\pi$ pulse then swaps the $\ket{1}-\ket{2}$ populations~\cite{Supplement}, enabling a second selective spill to imprint the $\ket{2}$ pattern via \textit{Mask}($\ket{2}$). This procedure grants deterministic access to the full local basis -- $\ket{\varnothing}$, $\ket{\uparrow}$, $\ket{\downarrow}$, $\ket{\updownarrows}$, enabling direct atom-by-atom construction of quantum states. The local and parallelized initialization of spin and density in the absence of spin-dependent optical potentials is the main result of this Letter. We demonstrate this control by preparing classical antiferromagnetic ordered states relevant for Fermi-Hubbard physics with engineered domain walls and precisely embedded defects [Fig.~\ref{fig:Image1_Teaser}(c)]. Our approach provides a highly programmable alternative for fermionic state preparation compared to the realization of band insulators and configurational derivatives thereof~\cite{Xu2025,Sompet2022,Bakr_2D}. The scheme benefits from the parallelized nature of deterministic local control, achieving arbitrary configurations in at most two spills. For the array sizes explored here, we observe no degradation in entropy per particle or ground-state fraction with increasing system size, consistent with the local character of the preparation stages and indicating the feasibility of larger arrays generated via AODs or spatial light modulators.

%%%%% \section{Conclusion and Outlook} \label{sec:conclusion}
In conclusion, we have demonstrated a novel approach for realizing arbitrary product states of fermions in optical tweezer arrays, utilizing spin-resolved deterministic preparation of $^6$Li atoms in the 3D motional ground state. These results establish microscopic control over the initial quantum state and enable quantum dynamical studies starting from programmable, low-entropy configurations of fermions. We showcase these capabilities by engineering target states of up to 128 atoms, within experimental cycle times of \qty{3}{\s}. For instance, a band-insulator-like pattern is realized on our platform with a $> $14\% success probability, corresponding to an average loading entropy of \qty{0.080 \pm 0.001}{\kb} per particle. Looking ahead, tunnel coupling such states in the tweezers or via transfer to optical lattices will enable the exploration of Fermi-Hubbard physics seeded with tailored features, such as hole pairs or spin-domain walls. At system sizes of $8\times8$, these characteristics place our platform in a regime comparable to that of quantum gas microscopes observing strong correlation phenomena at low temperatures \cite{bohrdt_review}, as well as to state-of-the-art experiments studying far-from-equilibrium dynamics \cite{endres_hse,aidelsburger_le}. They also match the system sizes accessible to leading numerical methods investigating charge and spin ordering in doped Hubbard systems \cite{blatz_stripeandpair, vnn_georges}. The array size in our implementation is limited by the diffusion dynamics during free-space fluorescence and the associated requirement for array expansion before imaging. Overcoming this limitation by techniques such as matter-wave magnification \cite{mwm_jochim, mwm_weitenberg} provides a direct route to scaling to larger system sizes. The preparation via singlets is compatible with further cooling via entropy redistribution~\cite{Xu2025} and postselection into the zero-magnetization sector~\cite{Bakr_2D}. More broadly, the capability to initialize fermionic quantum states with programmable spin composition offers a direct route to itinerant ferromagnetism and Fulde–Ferrell–Larkin–Ovchinnikov physics in spin-imbalanced systems~\cite{itinerantFM_MCproposal, fflo_review}.

\begin{acknowledgments}

\textit{Acknowledgments}---We thank the Heidelberg Quantum Architecture team at Heidelberg University for the initial development and continued collaboration on the modular quantum gas platform. We are grateful for helpful exchanges with the Lithium quantum gas microscope team and the FermiQP team at MPQ and thank Gaurav Vaidya, Cady Feng, Daniel Dux, Jonas Kruip, Xinyi Huang, Kirill Khoruzhii, Jun Ong, Pragya Sharma, along with Anton Mayer and Milan Antic, for their support during the initial stages and construction of the apparatus. We thank Adam Kaufman and Petar Bojović for a careful reading of the manuscript. This work was supported by the Max Planck Society, the European Union’s Horizon 2020 and Horizon Europe research and innovation programs (Grant Agreements No 948240 — ERC Starting Grant UniRand -- and No 101212809 -- ERC Proof of Concept FermiChem) and Germany's Excellence Strategy (EXC-2111-390814868).

\end{acknowledgments}
% %%%%%%%%%%%%%%%%%%%%%%%Bibliography%%%%%%%%%%%%%%%%%%%%

\providecommand{\noopsort}[1]{}\providecommand{\singleletter}[1]{#1}%

\clearpage
\appendix

\onecolumngrid 

\section{SUPPLEMENTARY INFORMATION} \label{appendix}
\setcounter{figure}{0}
\renewcommand{\thefigure}{S\arabic{figure}}
\renewcommand{\theHfigure}{S\arabic{figure}}

\makeatletter
\setcounter{NAT@ctr}{0}   
\setcounter{enumiv}{0}     

\renewcommand{\theNAT@ctr}{S\arabic{NAT@ctr}}
\renewcommand{\theenumiv}{S\arabic{enumiv}}
\makeatother

\subsection{Optical Traps and Potentials} \label{appendix:traps}

\subsubsection{Reservoir Preparation}
Our experiment utilizes a 2D-MOT to 3D-MOT loading scheme for obtaining cold atoms of $^6$Li at $\gtrsim $~\qty{e8}{atoms \per\s} flux~\cite{Li2D_MOT_supp, hqa_supp}. After loading the MOT for \qty{300}{\ms}, we load a far off-resonant \qty{1070}{\nm} crossed optical dipole trap (ODT) at \qty{60}{\W} in a nano-textured octagonal glass cell. The ODT has a crossing half-angle of \ang{10} between two \qty{67}{\um} waist beams in the $xy$ plane. The optically trapped $\ket{1} - \ket{2}$ mixture is balanced using an RF drive, yielding $\approx \,$10$^6$ atoms per spin. We perform forced evaporation at \qty{770}{\G} down to \qty{450}{\mW} where the trap frequencies are given by $\omega_x/2\pi \approx$ \qty{60}{\Hz}, $\omega_y/2\pi \approx \omega_z/2\pi \approx$ \qty{690}{\Hz} ($T_{\mathrm{F}} \approx$ \qty{628}{\nano\kelvin} with $\approx$ 1.3$\times$10$^4$ atoms per spin), to obtain a \qty{700}{\nano\kelvin} bulk gas at the threshold of quantum degeneracy. Ramping the magnetic fields quickly to \qty{300}{\G} at this point restrains Li$_2$ molecule formation, and establishes a two-component reservoir with $\approx \,$\qty{1.3e4}{atoms} per spin for tweezer loading.

The reference of our experimental setup is defined by the focal point of the 0.65 NA, FoV = \qty{\pm 100}{\um} objective (\textit{Special Optics}), that has been threaded onto a stationary composite assembly, monolithically attached to the optical breadboards. We leverage the mobile vacuum chamber design developed in collaboration with the Heidelberg Quantum Architecture group~\cite{hqa_supp}, and precisely position the glass cell obtaining a PSF Strehl ratio $>$0.95. Using a combination of long-pass dichroic mirrors, we employ the microscope to project \qty{808}{\nm} tweezers, and incoherent \qty{635}{\nm} light (\textit{Wavespectrum WSLS-635}) potentials via a digital micro-mirror device (DMD, \textit{Texas Instruments DLPLCR9000EVM}), while also collecting the \qty{671}{\nm} light for fluorescence detection (Fig.~1 in the main text).

\subsubsection{Tweezer Array Generation} \label{tw_array}

Far off-resonant $\lambda_{\mathrm{T}}$ = \qty{808}{\nm} light from a holographic grating stabilized 500 mW laser diode (\textit{Thorlabs LD808-SEV500}) is focused by the $f \,$=~\qty{18.8}{\mm} microscope objective to an array of optical tweezer traps. The array is generated by two crossed acousto-optic deflectors (AODs, \textit{Isomet OAD1343-XY-T70S-9}) that each provide more than 350 resolvable spots across a \qty{192}{\um} field-of-view given by the \qty{30}{MHz} AOD diffraction bandwidth. The microscope produces average tweezer waists of $w_0 \,$=~\qty{560 \pm 30}{\nm} across the 8$\times$8 array, falling within 10\% of the diffraction limit. The waists were measured using modulation spectroscopy at \qty{45}{\uW} per tweezer, yielding axial and radial trap frequencies $\omega_a/2\pi \approx \, \,$\qty{22}{\kHz} and $\omega_r/2\pi \approx \, \,$\qty{70}{\kHz}, respectively. 
For enhanced diffraction efficiency and telecentric operation, the \qty{808}{\nm} beam traverses a 4f-relay between the two AODs, resulting in total efficiencies of more than 70\% across the AOD bandwidth.
Radio-frequency tones of variable frequency and amplitude are simultaneously replayed from an arbitrary waveform generator at 625 million samples per second (\textit{Spectrum Instrumentation M4i.6621-x8}). This allows for full temporal control over tweezer position and trap depths utilizing $4N$ degrees of freedom (frequencies and amplitudes) on the AODs for an $N\! \! \times \! \! N$ array. We find it critical to ensure phase continuity of the RF tones between moves and static configurations of the array to avoid heating. Moreover, we minimize constructive interferences between the different tones such  that intermodulation is minimized at the spill configuration~\cite{rf_phase_supp}. 

We measure average motional ground-state lifetimes (1/e) of 17.5(1.0)~\unit{\s} after preparing singlet pairs, agreeing with estimates given by off-resonant scattering~\cite{GRIMM200095_supp}. The ground-state lifetime was measured by holding the optical tweezers after a singlet spill for variable times and performing a second spill right before imaging, wherein the motional excitations during the hold manifest as an atom loss. Moreover, for singlets, we observe lifetime corrected motional ground-state move fidelities of 99.8(0.1)\% per move at maximum average velocities of \qty{34}{\um/ \ms} when using minimum jerk trajectories~\cite{mjt_supp}. This agrees with the limits set by acoustic lensing inside the AOD crystals.  

\subsection{Deterministic Preparation with Gravity Spill} \label{Gravity Spill}

\begin{figure}
    \centering\
    \includegraphics[width=100mm]{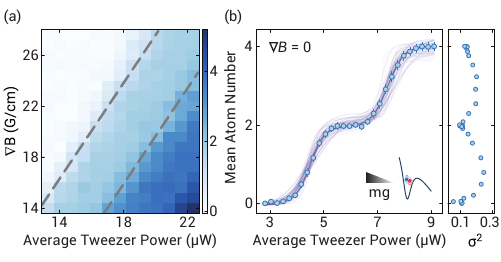}
    \caption{\textbf{Deterministic state preparation with varying potential gradients. (a)} Heatmap of the mean atom number as a function of tweezer power and applied magnetic field gradient. The dashed lines delineate the plateaus where even atom numbers can be deterministically prepared, following an approximately linear relation. \textbf{(b)} With gravity as the only potential slope, tweezer power ramps yield clear 2- and 4-atom plateaus. Light traces show individual tweezer spilling trends, following the 8$\times$8 array-averaged trend (dark trace). A double sigmoid fit separates the 0-2 and 2-4 transitions by $5 (\sigma_{0-2} + \sigma_{2-4})$, enabling robust spilling. The corresponding atom number variance is shown on the right.}
    \label{fig:gravity_spill}
\end{figure}

The deterministic few-fermion preparation scheme introduced in~\cite{fewfermionprep_supp}, leverages the degeneracy of a Fermi gas loaded into tightly confining optical traps, together with tunable optical power and magnetic field gradients, to control the number of bound states. This enables number-resolved preparation of fermions, as demonstrated in the main text. A key requirement for this scheme is precise control of trap depths with resolution better than the level spacing. Tightly confining traps make this feasible with percent-level power stabilization to produce well-resolved even-atom plateaus, reflecting the sequential filling of each bound state with two spin components. For $^6$Li, the large magnetic moment ($-\mu_{\mathrm{B}}$) at experimentally relevant fields allows the spill to be performed under a magnetic field gradient that defines a clean potential slope for atomic removal and enables stabilization at higher absolute trap powers than the ambient noise. In our setup with \qty{560 \pm 30}{\nm} tweezers, spill powers of few tens of \unit{\micro \W}($\approx \,$\qty{100}{kHz} trap depth) at $\nabla B \approx$\qty{20}{\G / \cm} yield plateau widths of $\approx\pm 5\%$. Scans over the magnetic gradient and tweezer power resolve the discrete level structure of the microtraps [Fig.~\ref{fig:gravity_spill}(a)].

We show that comparable control can be achieved without magnetic field gradients by using gravity as the potential slope. For $^{6}\mathrm{Li}$, gravity generates a potential gradient $mg \, \approx \, h \times$\qty{1.47}{\MHz / \cm} ($^{6}\mathrm{Li}$ mass $m$, acceleration due to gravity $g$, and Planck's constant $h$), equivalent to a magnetic field gradient of \qty{1.1}{\G / \cm}. Extrapolating the linear shift of the even-atom plateaus as in Fig.~\ref{fig:gravity_spill}(a) to this value agrees with the observed trap depths required to unbind the corresponding excited states [Fig.~\ref{fig:gravity_spill}(b)]. Because the gravitational potential scales with mass, and $^{6}\mathrm{Li}$ is among the lightest fermionic species used in atomic arrays, this measurement can be considered to serve as a lower bound for relative power stability needed for robust spilling in a given trap geometry. In our case, average powers of $\approx \,$\qty{6 \pm 0.3}{\uW} per tweezer yield gravity-induced spilling with single-particle preparation fidelities of \qty{98.7 \pm 0.1}{\%} averaged over an 8$\times$8 array, at the same offset field (\qty{527}{\G}) used for gradient-assisted spilling in the main text. These results indicate that gravity-based spilling may offer a practical route to deterministic preparation for atomic species such as $^{87}\mathrm{Sr}$ and $^{171}\mathrm{Yb}$, which are relevant for quantum simulation and information processing applications, but have magnetic moments on the order of $\approx \! 10^{-4} \mu_{\mathrm{B}}$~\cite{akali_earth_magnetic_moment_supp}. For such systems, harnessing fermionic statistics together with gravity could provide an attractive pathway for reliable initialization of ground-state fermionic modes in optical tweezers.

\subsection{Antenna Design and Characterization} \label{appendix RF-MW}

High-fidelity radio-frequency (RF) and microwave (MW) transitions with strong, controllable Rabi couplings are essential for our single-exposure spin-resolved imaging scheme. Building on earlier antenna implementations~\cite{Roati_antenna_supp}, we develop compact, modular designs optimized for $^6$Li, achieving substantially enhanced current transfer efficiencies and coupling strength. These antennas enable the highest Rabi rates reported for the relevant hyperfine transitions, extending the range of accessible state manipulation protocols in $^6$Li systems.

\begin{figure*}[ht]
    \centering\
    \includegraphics[width=170mm]{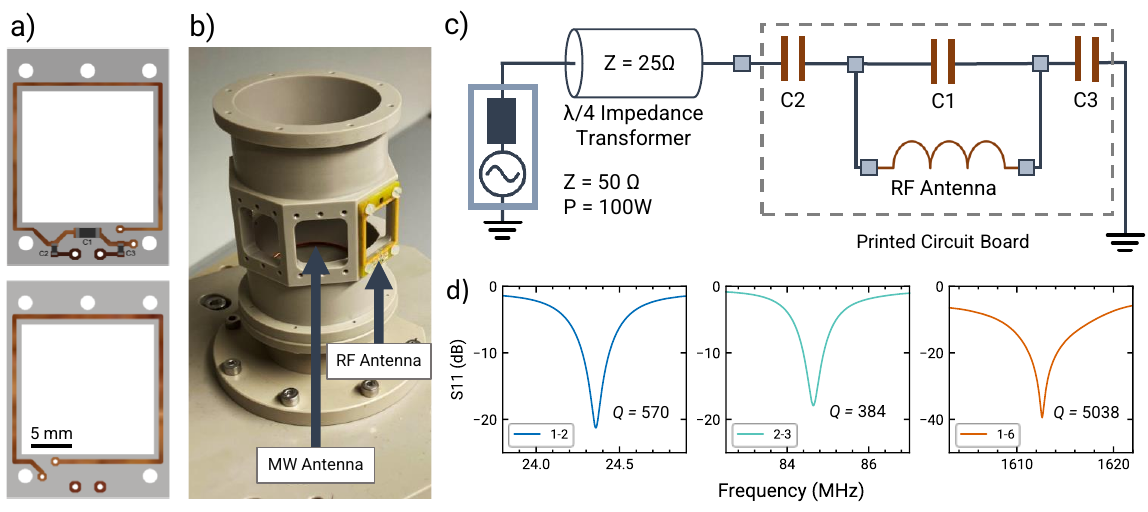}
    \caption{\textbf{Design and realization of RF and MW antennas. (a)} PCB diagram of the $\ket{2} - \ket{3}$ RF antenna. The top figure shows the front surface of the PCB while the bottom figure depicts the back-side which faces the atoms. Capacitor $C1$ is SMD case 1808 and $C2$, $C3$ are 0603 ($C1$ uses TDK Multilayer Ceramic Capacitor, with voltage thresholds up to \qty{3}{\kV} DC, while $C2$ and $C3$ may be general purpose $\approx \,$\qty{100}{\V} threshold capacitors). Front and back surfaces are connected via through-holes. The signal carrying coax cable is directly soldered onto the board through the two holes at the PCB bottom. \textbf{(b)} We use an octagonal design for a PEEK cage surrounding the nano-texture coated octagonal glass cell to mount antennas and bias coils close to the atoms. Here the mounted RF and MW antennas are shown in a test setup.\textbf{(c)} Circuit diagram for the RF antenna: a $25 \Omega$ $\lambda/4$ length coax cable transforms the impedance from $50\Omega$ to $12.5\Omega$ as proposed in~\cite{Roati_antenna_supp}. Capacitor values are chosen to impedance-match the circuit at the desired frequency. \textbf{(d)} $S11$ parameter for the RF and MW antennas, measured on a vector network analyzer. $Q$-factor is calculated as the ratio of center frequency to \qty{-3}{dB} width.}
    \label{fig:Image_Antenna_v2}
\end{figure*}
\subsubsection{Radio Frequency Antennas} \label{appendix RF}
Detection of state $\ket{2}$ proceeds via population transfer to $\ket{3}$ prior to imaging. To drive this transition, we use a compact printed circuit board (PCB) loop antenna placed $\approx \,$\qty{30}{\mm} from the atoms to maximize near-field coupling. The antenna, shown in Fig.~\ref{fig:Image_Antenna_v2}(a), consists of two square loops (side length L=\qty{20}{\mm}) made from \qty{1}{oz} (\qty{35}{\um} thickness) copper wire. It is driven by a \qty{100}{\W} RF amplifier (\textit{Mini-Circuits ZHL-100W-GAN+}), with duty cycles $<$\qty{10}{\ms} every \qty{3}{\s} to prevent PCB heating under high power-transfer efficiency. The antenna axis is orthogonal to the atomic quantization axis, conducive for driving the desired $\Delta m_I = 1$ nuclear-flip transitions.

At the \qty{527}{\G} field used during detection, the $\ket{2}-\ket{3}$ transition lies at \qty{84.6}{\MHz}, with a magnetic dipole matrix element $\bra{2}\mu_x\ket{3}/h\approx$ \qty{107}{\kHz / \G}. The antenna is impedance-matched to resonate at this frequency, yielding Rabi oscillations at $2\pi \times$\qty{66}{\kHz}, corresponding to magnetic field peak amplitudes of $\approx \,$\qty{620}{\milli \G} at the atoms. 

Because the antenna dimensions are much smaller than the RF wavelength, radiation losses can be neglected and loops can be viewed as coils with a characteristic impedance. Frequency tuning is achieved through appropriate capacitor selection, allowing straightforward adaptation of the same design for addressing other transitions. For instance, to drive the $\ket{1}-\ket{2}$ transition at \qty{27.1}{\G} for the spin-selective preparation scheme, we use a four-loop variant of identical geometry and match it to the \qty{24.4}{\MHz} resonance using off-the-shelf capacitors. Driven by a \qty{5}{W} amplifier (\textit{Mini-Circuits ZHL-5W-1+}), this antenna realizes a Rabi frequency of $2\pi \times$\qty{166}{\kHz}, implying magnetic field peak amplitudes of \qty{364}{\milli \G} at the atoms, given the transition's \qty{456}{\kHz / \G} matrix element.

Both RF antennas exhibit \qty{-3}{\dB} bandwidths of order \qty{100}{\kHz}, corresponding to a $Q$-factor exceeding 300 [see Fig.~\ref{fig:Image_Antenna_v2}(d)]. This is possible due to the antenna and the tuning capacitor ($C1$) forming an LC-resonator, in combination with a $\lambda/4$ impedance transformer~\cite{Roati_antenna_supp} that enables substantial current buildup relative to the amplifier output. These high-$Q$ resonances enable strong atomic coupling, exemplified by the coherent Rabi oscillations, and discrete $\pi$-pulses yielding $> \,$99.5\% per-pulse fidelity for more than 45 repetitions. For operation at high duty cycles or with continuous drive, the performance can be further improved using higher-$Q$ capacitors, thicker copper traces, and enhanced thermal management. 

\subsubsection{Microwave Antenna} \label{appendix MW}
For imaging atoms in state $\ket{1}$, we transfer the population to the stretched state $\ket{6}$ using a resonant MW transition at \qty{1.637}{\GHz} and \qty{527}{G}. The transition has a coupling strength of \qty{1.39}{\MHz / \G}. We drive this using a minimal single-loop MW antenna made with \qty{2}{\mm^2} cross-section copper wire into a \qty{16.6}{\cm} circumference (roughly one free-space wavelength), and positioned $\approx \,$\qty{3}{\cm} from the atoms [see Fig.\ref{fig:Image_Antenna_v2}(b)]. The loop is directly soldered to the end of a \qty{50}{\Omega} low-loss SMA cable (\textit{Qualwave QA800-5.6}) offering high phase-stability, without additional matching circuitry. The simplicity of this geometry minimizes losses at microwave frequencies and provides convenient experimental access. 

Driven by a \qty{25}{\W} MW amplifier (\textit{Mini-Circuits ZHL-25W-272X+}), the antenna achieves a Rabi frequency of $2\pi \times $\qty{80}{\kHz}, corresponding to a peak magnetic field amplitude of $\approx \,$\qty{60}{\milli\G} at the atoms. Owing to the large differential magnetic moment, the transition is sensitive to magnetic field noise at the \qty{10}{\milli\G} level, and to power fluctuations in the MW drive, which are not actively stabilized in the present setup. To ensure repeatable, high-fidelity population transfer under these conditions, we use an I/Q modulated GOIA-WURST (Gradient Offset Independent Adiabatic pulse with WURST modulation) waveform generated by a MW arbitrary waveform generator~\cite{GOIA-WURST_supp}. This composite pulse combines the adiabatic frequency sweep of a GOIA pulse with the smooth amplitude envelope of a WURST modulation, a technique widely used in the context of nuclear magnetic resonance spectroscopy for noisy environments. With this approach, we achieve transfer fidelities exceeding 99\% for a single \qty{15}{\us}~$\pi$-pulse.

\subsection{Imaging Scheme and Single Atom Detection} \label{appendix:singleatomresolution}

\begin{figure}
	\centering\
	\includegraphics[width=100mm]{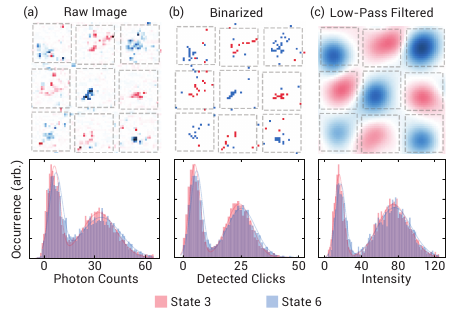}
	\caption{\textbf{Image Processing for Single-Atom Resolution. (a)} A typical single-exposure raw camera image showing overlapped spin sectors (top row). Bottom: photon-count histograms for $\ket{3}$ (Red) and $\ket{6}$ (Blue), compiled from $\approx \,$8000 single atom realizations. Dashed boxes indicate candidate sites defined by known tweezer positions. The 0- and 1- atom discrimination fidelity is 95.0(1)\%. \textbf{(b)} Binarization using an optimized intensity threshold converts pixel intensities into discrete camera clicks, improving discrimination fidelity to 98.5(1)\%. \textbf{(c)} Applying a low-pass filter to the binarized image enhances photon clustering information from atomic fluorescence, in contrast to uncorrelated noise clicks, yielding a final fidelity exceeding 99.4(1)\% for both spins.}
	\label{fig:NuvuFidelity}
\end{figure}

The spin-resolved imaging introduced in this work conceptually builds upon the free-space imaging and detection framework developed for $^6$Li in earlier works~\cite{bergschneider1_supp, Su2025_supp}, but enables parallel single-exposure readout (also see~\cite{HQA_imaging_supp}). The orthogonal $\sigma^{\pm}$ fluorescence components that encode the spin information are optically separated onto distinct regions of the camera (see Fig.~2 in the main text). To facilitate direct comparison, the two spin sectors can be laterally overlapped in post-processing by combining and subtracting the corresponding regions of interest as in Fig.~1 and Fig.~\ref{fig:NuvuFidelity}, whereas the rest of the image processing strategies remain the same as elucidated in the following. 

The EMCCD camera (\textit{Nüvü HNü 512} $\gamma$) allows for high fidelity discrimination between zero and one photon events. Multi-photon events cannot be resolved due to the stochastic nature of the gain process. We therefore employ an imaging path that reduces the probability of multiple photons striking the same pixel, thus leveraging the single-photon counting mode of the EMCCD. The camera has \qty{16}{\um} pixel size, corresponding to \qty{1.5}{\um} in the atom plane at a magnification of $\approx \,$10.5. During the \qty{19}{\us} exposure, atoms undergo a diffusive random walk, averaging to a Gaussian $1/e$ waist of \qty{6.6 \pm 0.2}{\um}, corresponding to an area of about 9$\times$9 pixels. Given the probe intensity of $5 I_{\text{sat}}$ and the $D2-$linewidth $\Gamma_{D2} = 2\pi \times \,$\qty{5.87}{\per \us}, each atom scatters approximately 290 photons at a rate of \qty{15.4}{photons /\us} during the imaging sequence, of which we detect $\approx \,$35 photons. This is consistent with the expected efficiencies: 16\% collection by the objective for the targeted $\sigma$-transitions accounting for the dipole emission pattern, 85\% optical transmission through setup, and 90\% EMCCD quantum efficiency at \qty{671}{\nm}, giving an overall detection efficiency of 12\%.
During imaging, all traps are turned off, and a \qty{671}{\nm} band-pass filter suppresses off-resonant background light. The residual background level originates mainly from clock-induced charges (CICs) in the EMCCD, and spurious scattering of the imaging beams. The dark CIC noise level of 0.3\%, rises to 0.6\% in presence of the resonant beams, however, the atomic fluorescence signal ($\approx \,$\qty{1}{photon/pixel}) exceeds the background by two orders of magnitude.

Image processing follows a photon-counting approach optimized for low-noise detection, given the raw histograms. Each frame is first processed with respect to a binarization threshold of $I_{\mathrm{bin}} = 5\sigma_{\mathrm{readout}}$, truly rejecting 99.7\% of dark pixels while retaining 82\% of real photon events~\cite{bergschneider1_supp, Su2025_supp}. A Gaussian low-pass filter then enhances spatially correlated photon clusters from atomic fluorescence relative to uncorrelated noise. This approach substantially improves discrimination fidelity, as shown in Fig.~\ref{fig:NuvuFidelity}. Candidate sites are identified within regions corresponding to the known tweezer positions, and low-passed intensity thus generated is compared to a calibrated discrimination threshold, for each site to be classified as occupied or empty. This procedure yields bimodal histograms with well-separated zero- and one-atom distributions, quantified by $>$99.4\% fidelities defined at the equal false positive and false negative probabilities threshold, as shown in Fig.~\ref{fig:NuvuFidelity}.

\subsection{Number-Resolved Detection} \label{Appendix:uMOT_counting}

At the photon numbers we work with, the Poisson distributions in photon counts corresponding to one or multiple atomic scatterers are not separated significantly~\cite{Su2025_supp}. This fundamentally limits our ability to distinguish between one- and many-atom occupations within one tweezer based solely on fluorescence amplitudes from an EMCCD camera. To ensure that our measured preparation fidelities are not compromised by this limited discrimination capability, we compare the EMCCD-based detection against a calibrated atom-counting method using high-gradient 3D magneto-optical trap (\textmu MOT).

\begin{figure}
    \centering\
    \includegraphics[width=82mm]{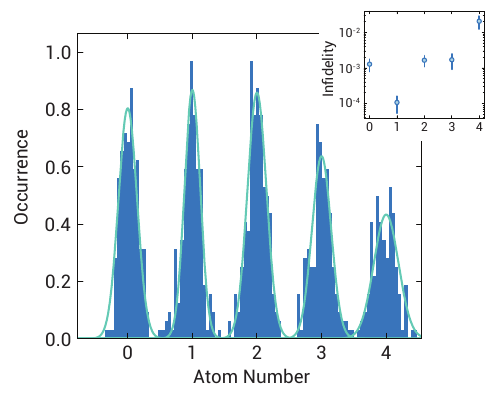}
    \caption{\textbf{Atom-resolved counting in a \textmu MOT} Fluorescence histogram from a compressed \textmu MOT show well-resolved atom number peaks, allowing reliable spin-agnostic benchmarking. \textit{Inset}: Gaussian fits yield discrimination fidelities exceeding 99.8\%.}
    \label{fig:single_atom_counting}
\end{figure}

Atom-resolved counting in the \textmu MOT follows established techniques introduced in~\cite{fewfermionprep_supp, motcounting_oberthaler_supp}. After the main experimental sequence, the ensemble is transferred to a large-volume MOT at a moderate magnetic field gradient of $\approx \,$\qty{30}{\G / \cm}, with trapping beams red-detuned by $\approx 4 \Gamma_{D2}$ from resonance. The MOT is then compressed by ramping the gradient to \qty{250}{\G / \cm}, and reducing the detuning to $\approx 1 \Gamma_{D2}$, forming a \qty{250}{\um}-scale \textmu MOT. The fluorescence emitted during a \qty{500}{\ms} exposure is collected on a CMOS camera (\textit{Alvium G1-240m}), producing well-separated peaks in the collected intensity histograms, corresponding to integer atom numbers. As shown in Fig.~\ref{fig:single_atom_counting}, this technique enables clear atom number discrimination, providing a benchmark for EMCCD-based measurements.

To compare the two methods, we perform the identical singlet preparation protocol as in the main text, with the exception that during the spilling stage, all tweezers except one are emptied using the site-selective DMD blowout procedure. The remaining tweezer is then analyzed using the \textmu MOT counting method, which allows a spin-agnostic exact atom-number measurement. Repeating this process across all sites, we acquire $\approx \,$150 realizations per tweezer over extended data runs to obtain statistically significant comparisons. The measured preparation fidelities from EMCCD imaging agree with the \textmu MOT counting results within experimental uncertainties, confirming that single-shot EMCCD detection faithfully reproduces the underlying occupation statistics. Furthermore, the \textmu MOT data shows that nearly all multi-atom events correspond to exactly two-atoms (thus, singlets), with over 99.2\% of the events containing $\geq2$ atoms being true two-atom pairs. This demonstrates the residual imperfections in the preparation originate dominantly from single-atom losses rather than higher-mode occupancies. Taken together, these benchmarking results validate the use of the EMCCD-based detection method to realize the rapid, parallelized single-exposure detection, and extracting preparation fidelities from the measurements therein.

\subsection{Ground State Fraction Estimation} \label{ground_state_fraction}

Obtaining high 3D ground state fractions with deterministic preparation is a challenge across neutral atom platforms, where sophisticated in-trap cooling schemes and light-shift measurements are employed for their initialization and characterization~\cite{SQM_Tweezers_supp, kaufman_quantum_walk_supp}. For our experiments, it is essential that the deterministically prepared $\ket{1}- \ket{2}$ singlet pairs occupy the motional ground states of each tweezer, for these to serve as an assembly resource for Hubbard quantum simulation.

Although the singlet preparation sequence is designed to initialize atoms in the 3D ground state, motional excitations may arise from several heating mechanisms, including non-adiabatic potential ramps, parametric heating from laser noise, residual trap motion from RF-tone interferences in the AOD, and background gas collisions. To verify that these effects are negligible, we perform a second spill after the completion of the experimental sequence. If excitations were present, this additional spilling step would remove atoms occupying higher motional states, resulting in a measurable loss in the mean atom number~\cite{fewfermionprep_supp}.

To minimize heating throughout the sequence, we optimize multiple aspects of the experimental protocol: suppressing relative intensity noise on the tweezer laser, implementing minimum-jerk trajectories for array translations, and applying Blackman-shaped power ramps during spilling with average slew rates of $\approx \,$\qty{0.7}{\uW / \ms}. These measures ensure adiabaticity and reduce motional excitations. 

From such additional spill measurements, we determine the probability of maintaining singlets in the motional ground state to be \num{100(0.0:0.2)}\%. Specifically, we perform a second spill after the initial spill in the preparation step and record additional losses exclusively due the second spill. The same verification procedure -- an additional spill following the intended sequence -- is used to benchmark ground-state fidelities throughout this work, including tweezer lifetime measurements, where a variable hold time is inserted between the two spills. Given the high precision and reproducibility of the spilling process, this approach provides a direct and reliable method for estimating 3D ground state populations in our $^6$Li arrays. 

\subsection{Entropy Estimation}
The entropy of the prepared product state originates dominantly from stochastic particle loss during the spilling process, assuming that the occupation of the ground state is initially near unity. This results in an average single-particle preparation fidelity of $p=\mathrm{98.47(2)\%}$ across the 8$\times$8 array ($p=\mathrm{98.7(1)\%}$ for gravity spill). Assuming independent preparation errors and equal probabilities $p$ to occupy a single-particle mode for both spins, the entropy per particle is 

\begin{equation*}
\frac{S}{\left<N\right>} = -\frac{\mathrm{k}_{\mathrm{B}}}{p} \left( p \ln{p} + \left(1-p\right) \ln{\left( 1-p \right)} \right),
\end{equation*}

where $\mathrm{k}_\mathrm{B}$ is the Boltzmann constant. The probability of preparing a singlet pair is $p_{\uparrow\downarrow}=p^2$, a hole $p_{\varnothing}=(1-p)^2$, and only a single atom $p_\uparrow=p_\downarrow=p(1-p)$. This yields an entropy per particle of $S/\left<N\right>$= \qty{0.080 \pm 0.001}{\kb} for a uniformly filled array (\qty{0.070 \pm 0.004}{\kb} for gravity-assisted spill). For patterned product states, such as the antiferromagnetic configuration, additional entropy arises from finite infidelity in the spin-selective removal step for retention of $\ket{2}$ atoms in the ground state of the addressed tweezers (see Fig.~4d in the main text). Assuming each tweezer is addressed once, and accounting for the $\ket{1}- \ket{2}$ RF-flip fidelity, this results in an increased entropy per particle of $S/\left<N\right>$= \qty{0.152 \pm 0.001}{\kb} (\qty{0.145 \pm 0.004}{\kb} for gravity spill). 

In optical lattices, band insulator preparation was achieved with $S/\left<N\right>$= \qty{0.025 \pm 0.004}{\kb}~\cite{Xu2025_supp}, with other experiments reaching typical values around $S/\left<N\right>$= \qty{0.3}{\kb} at half filling~\cite{Sompet2022_supp}. Experiments with tunnel-coupled tweezers have demonstrated $S/\left<N\right>$= \qty{0.12 \pm 0.03}{\kb} in the independent loading stage for a 1$\times$8 array~\cite{Bakr_1D_supp}, and $S/\left<N\right>$= \qty{0.34 \pm 0.01}{\kb} in 5$\times$5 configurations~\cite{Bakr_2D_supp}. 
These values benchmark the low-entropy initialization achievable with our  preparation scheme and its future use for quantum gas assembly in optical lattices.

% \bibliography{Bibliography}
\providecommand{\noopsort}[1]{}\providecommand{\singleletter}[1]{#1}%

\end{document}